\documentclass[aip,reprint,amsmath,amssymb]{revtex4-1}

\usepackage{graphicx}
\usepackage{textcomp}

\begin{document}


\title{High quality factor mg-scale silicon mechanical resonators for 3-mode optoacoustic parametric amplifiers}


\author{F. A. Torres}
\email[]{francis.achilles.torres@physics.uwa.edu.au}
\affiliation{School of Physics, University of Western Australia, 35 Stirling Highway, Crawley, Western Australia 6009, Australia}

\author{P. Meng}
\affiliation{School of Physics, University of Western Australia, 35 Stirling Highway, Crawley, Western Australia 6009, Australia}

\author{L. Ju} 
\affiliation{School of Physics, University of Western Australia, 35 Stirling Highway, Crawley, Western Australia 6009, Australia}

\author{C. Zhao}
\affiliation{School of Physics, University of Western Australia, 35 Stirling Highway, Crawley, Western Australia 6009, Australia}

\author{D, G. Blair} 
\affiliation{School of Physics, University of Western Australia, 35 Stirling Highway, Crawley, Western Australia 6009, Australia}

\author{K.-Y. Liu}
\affiliation{Australian National Fabricatoin Facility (Queensland Node), Australian Institute for Bioengineering and Nanotechnology, The University of Queensland, Brisbane, QLD 4072, Australia}

\author{S. Chao}
\affiliation{Institute of Photonics Technologies and E.E. Dept., National Tsing Hua University, 101 Kuangfu Rd. Sec. 2, Hsinchu, Taiwan 300}

\author{M. Martyniuk}
\affiliation{School of Electrical and Electronic Engineering, University of Western Australia, 35 Stirling Highway, Crawley, Western Australia 6009, Australia}

\author{I. Roch-Jeune}
\affiliation{Plate-Forme RENATECH IEMN, Ave Poincar$\acute{e}$ - BP 60069, 50652 Villeneuve d'Ascq cedex, France}

\author{R. Flaminio}
\affiliation{Laboratoire des Mat$\acute{e}$riaux Avanc$\acute{e}$s (LMA), IN2P3/CNRS, Universit$\acute{e}$ de Lyon, F-69622 Villeurbanne, Lyon, France}

\author{C. Michel}
\affiliation{Laboratoire des Mat$\acute{e}$riaux Avanc$\acute{e}$s (LMA), IN2P3/CNRS, Universit$\acute{e}$ de Lyon, F-69622 Villeurbanne, Lyon, France}

\date{\today}

\begin{abstract}

Milligram-scale resonators have been shown to be suitable for the creation of 3-mode optoacoustic parametric amplifiers, based on a phenomena first predicted for advanced gravitational-wave detectors. To achieve practical optoacoustic parametric amplification, high quality factor resonators are required. We present millimetre-scale silicon resonators designed to exhibit a torsional vibration mode with a frequency in the $10^{5} - 10^{6}$ Hz range, for observation of 3-mode optoacoustic interactions in a compact table-top system. Our design incorporates an isolation stage and minimizes the acoustic loss from optical coating. We observe a quality factor of 7.5 \texttimes $10^{5}$ for a mode frequency of 401.5 kHz, at room temperature and pressure of $10^{-3}$ Pa. We confirmed the mode shape by mapping the amplitude response across the resonator and comparing to finite element modelling. This study contributes towards the development of 3-mode optoacoustic parametric amplifiers for use in novel high-sensitivity signal transducers and quantum measurement experiments.
\end{abstract}


\maketitle

\section{Introduction}
%

Three-mode optoacoustic parametric interactions were first introduced by Braginsky \textit{et al}. \cite{Braginsky2001,Braginsky2002} who presented the phenomena and evaluated the potential risk of 3-mode parametric instability in long optical cavities for gravitational-wave detectors \cite{Abbott2009}.

The physics of the 3-mode optoacoustic parametric amplifier is similar to that of an optical parametric amplifier, in which the interaction occurs through the Kerr effect. The 3-mode optoacoustic interaction occurs when 2 optical modes inside an optical cavity are resonantly coupled to a mechanical mode of the cavity mirror. The coupling is mediated by the radiation pressure forces due to the beating of the two optical modes. Three-mode interactions have been studied extensively \cite{Miller2011,Vyatchanin2012}, and observed, in long optical cavities \cite{Zhao2008}. 

As triply resonant devices, 3-mode optoacoustic parametric amplifiers can in principle have very high sensitivity to torsional motion, achieve strong optoacoustic coupling and require less laser power than a 2-mode interaction \cite{Abbott2009,Cohadon1999}, where only 2 modes are resonant. For this reason, the 3-mode interaction reduces susceptibility to laser phase noise and amplitude noise \cite{Zhao2011}.

The creation of 3-mode parametric amplifiers requires a cavity design in which the mechanical motion of the mirror couples the main laser carrier mode to a transverse mode. For example, an optimal optoacoustic coupling can be achieved by using a torsionally resonant mirror, which has optimal spatial overlap to the TEM$_{01}$ optical cavity mode. The mechanical mode of frequency near the MHz range can be matched to the gap between the TEM$_{00}$ and TEM$_{01}$ modes in a specially designed optical cavity, as discussed by Miao \textit{et al.} \cite{Miao2009}. This paper describes mechanical resonators designed for table-top 3-mode parametric amplifiers \cite{Zhao2009}.

In a small optical cavity, 3-mode optoacoustic parametric instability was first observed by Chen \textit{et al.} with a silicon nitride membrane of thickness 50 nm, within a Fabry-P$\acute{e}$rot cavity \cite{Chen2013}. Miao \textit{et al.} \cite{Miao2009} have proposed a tunable compact table-top system, designed to observe 3-mode optoacoustic parametric interactions by adding a tuning cavity to a main optical cavity. This would allow continuous tuning between the positive gain regime (characterized by the Stokes mode, or amplification of the acoustic mode) and the negative regime (Anti-Stokes mode, referred to as `self-cooling' in the literature \cite{Kippenberg2008}), by small adjustments of a lens and mirror.

%
%

In order to produce 3-mode optoacoustic parametric amplifiers in a compact table-top setup, a mechanical resonator is required to be in the mm- and mg-scale ranges, have a high mechanical quality factor, and a mechanical oscillation near the 0.1 to 1 MHz range.

\begin{table}[b]
\centering
\begin{tabular}{|r|r|r|r|r|}

       \hline
       Authors & Mass & Mode Shape & Frequency &Q(300K)       \\ \hline
Davis \textit{et al.}\cite{Davis2010}&0.1 pg&Torsion & 21 MHz &2000\\ \hline
Chabot \textit{et al.}\cite{Chabot2005}&0.7 ng &Torsion&120 kHz&12000 \\ \hline
Arcizet \textit{et al.}\cite{Arcizet2008}&210 $\mu$g&(0,4) mode&2.8 MHz&15000\\ \hline
Serra \textit{et al.}\cite{Serra2012}&3 mg&Torsion & 85.5 kHz &145000 \\ \hline
Kuhn \textit{et al.}\cite{Kuhn2011}&1 pg&Longitudinal&3.66 MHz&1.8\texttimes $10^{6}$ \\ \hline
         
        \hline

\end{tabular}
\caption{Review of mechanical resonators reported in the literature with high quality factors and frequencies in the range of interest.}
\label{tab:ReviewResonators}
\end{table}

Mechanical resonators have been reported in the literature and have characteristics which are comparable to these requirements. Nanomechanical resonators were studied by Davis \textit{et al.} \cite{Davis2010}, and a quality factor of 2000 was reported for a torsional mode with effective mass of 0.1 pg at frequency 21 MHz. Measurements were performed at room temperature and pressure of $\sim 10^{-7}$ Pa. A pattern of 3 `paddles' (rectangular elements) along a rod of width 100 nm was used to isolate the vibration of the central paddle from vibrations of the surrounding silicon nitride membrane. This relatively low quality factor could be attributed to acoustic coating losses, as the entire sample was coated on one side with 10 nm of permalloy, for use as a torque magnetometer.

%
Similar nanomechanical resonators were fabricated and studied by Chabot \textit{et al.} \cite{Chabot2005}. They achieved mechanical quality factors of 12 000 for torsional modes at 120 kHz, with effective mass of 0.7 ng, measured at room temperature and pressure of 13 Pa. These resonators were made by etching silicon (100) boron-doped wafers, resulting in multiple double-torsion paddle designs. A small magnetic film dot of 3 $\mu$m diameter was coated onto the upper paddle, in view of making magnetometers and force sensors in nuclear magnetic resonance force microscopy.


A French team led by Arcizet \textit{et al.} fabricated and studied mm-scale silicon-on-insulator chip resonators \cite{Arcizet2008}, made using double-sided photolithography and deep reactive-ion etching (DRIE). They reported quality factors of 15 000, for a (0,4) transverse mode of 2.8 MHz, with effective mass of 210 $\mu$g, measured at room temperature and pressure of 0.1 Pa. These 1 mm by 1 mm beam resonators were optically coated for high-sensitivity optical monitoring of moving micromirrors.


Recently, Serra \textit{et al.} reported \cite{Serra2012} quality factors as high as 1.5 \texttimes $10^{5}$ for torsional modes of a mm-scale central paddle micromechanical silicon resonator, resonating at 85.5 kHz at room temperature and pressure of $10^{-3}$ Pa. Vibration isolation paddles and varying the thickness of different parts of the sample were used to shield the main resonating paddle mode from wafer modes. The aim of their design was to generate non-classical states of light by opto-mechanical coupling, and to produce devices suitable for the production of pondero-motive squeezing, and entanglement between macroscopic objects and light.


%
Kuhn \textit{et al.} \cite{Kuhn2011} made and studied 1 mm long, 240 $\mu$m wide, triangular nanopillars made of crystaline quartz (known for low loss). They obtained a quality factor of 1.8 \texttimes $10^{6}$ at 3.66 MHz, for a longitudinal mode of the pillar, measured at $10^{-1}$ Pa at room temperature. Table \ref{tab:ReviewResonators} summarises the results obtained by the above mentionned researchers.

%

For various reasons, the resonators presented above are not directly useful to use as 3-mode optoacoustic parametric amplifiers. However, they have helped to inspire our design, presented in the next section.


The above mentionned mm-scale resonators can be significantly outperformed
by larger bulk silicon resonators. Results were obtained for boron-doped bulk silicon cylindrical resonators with length between 6 to 75 mm, and common diameter of 76.2 mm. Nawrodt \textit{et al.} \cite{Nawrodt2008} report a quality factor as high as $\sim$ 3.5 \texttimes $10^{6}$. This corresponds to the fundamental drum mode of a cylindrical sample of length 12 mm, for a mode frequency $\sim$ 14 kHz at room temperature and pressure of $10^{-3}$ Pa. The highest quality factor obtained for the same sample, measured at 5.6K, was 4.5 \texttimes $10^{8}$.

%
%

Most resonators have quality factors far below what is obtained for bulk silicon resonators. However, this paper presents resonators that have quality factors that do come close.

Our motivation, comparable to the goals of Serra \textit{et al.} \cite{Serra2012}, was to test 3-mode optoacoustic parametric amplifiers in table-top setups, that are capable of being tuned between positive gain and self-cooling \cite{Cohadon1999}.

%
%

In this paper, we report our design and finite element modelling (FEM) of millimetre-scale silicon mechanical resonators, and various methods of fabrication. We describe the experimental setup and methods used for measurements. We report our results, which are getting close to the low end of quality factors for bulk resonators \cite{Nawrodt2008}, and a substantial quality factor improvement from mm-scale resonators reported above.

\section{Resonator Design, Modelling and Fabrication}

The primary purpose of our design, inspired by previous work with a silicon nitride membrane \cite{Davis2010}, was to achieve a torsional mode that has adequate vibration isolation, a frequency $\sim$ 400 kHz, and an adequate spatial overlap with a supported optical transverse mode: a requirement for 3-mode optoacoustic parametric interactions \cite{Zhao2009, Torres2010}. For reasons of simplicity and low cost of fabrication, we avoided designs that required step changes in thickness. Silicon was chosen for its low cost, availability, ease of fabrication and low acoustic loss \cite{Nawrodt2008, Arcizet2008, Davis2010}.
It is important to note that while a good vibration isolation is required to observe mechanical modes of interest with high quality factor, too much isolation is also a problem, if we want to excite the mode of interest via a different part of the resonator which is not directly on the torsional paddle itself. A balance must be reached to have low enough isolation to allow excitation of the central paddle mode, via piezo excitation of the wafer, and high enough to observe a high quality factor of that mode.

\begin{figure}[t]
    \centering
    \includegraphics[width=3.4in]{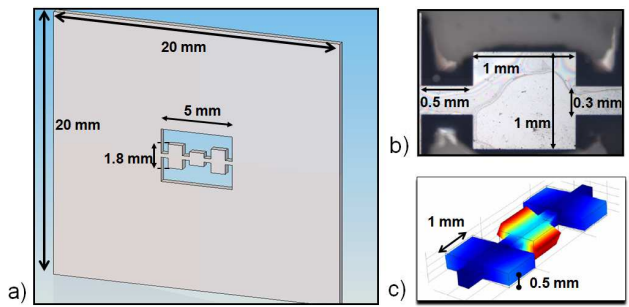}
    \caption{ (color online). Micro-mechanical resonator. (a) Wafer design: a 20 mm by 20 mm wafer of thickness 500 $\mu$m, with a pattern of three paddles on a 5 mm long torsion rod. (b) Optical microscope image of the central element of the resonator: a square mm paddle connected by torsion rods of length 0.5 mm and width 0.3 mm. (c) Finite element modelling view of the 3 paddles, each with a width of 1 mm and thickness 500 $\mu$m. Side paddles have a length of 1.8 mm. This design was chosen to obtain a torsional mode with high frequency and a central paddle to act as a rigid body.}
    \label{Our_Resonators}
\end{figure}

Our design consists of a 20 mm by 20 mm wafer with a central 1 mm by 1 mm paddle and two side paddles (1 mm by 1.8 mm each). These paddles are on a narrow rod of length 5 mm and width of 0.3 mm (see Fig. \ref{Our_Resonators}). The entire sample, made of silicon (100) boron-doped monocrystal, has a uniform thickness of 500 $\mu$m.

%
The central paddle is designed to undergo minimal elastic deformation and act as a rigid body. The elastic stress is thus confined to the torsional rods, and an area of minimal elastic stress of 0.8 mm by 0.8 mm, centered on the central paddle, is reserved for optical coating \cite{Zhao2009}. This is important to reduce the loss due to stress on the coating material, leading to unwanted contributions to the acoustic loss.

%
%

%

Simulations by FEM, using ANSYS 14.0 and a mesh size of 200 $\mu$m for the wafer, and 50 $\mu$m for the central paddle, showed that the fundamental wafer drum mode appeared around 10.64 kHz, while higher frequency modes were mostly paddle vibration modes. Modelling indicated that the 3-paddle system should exhibit 3 torsion modes, of which the mode with highest quality factor (mode of interest) would have most vibrations localised in the central paddle. In particular, the torsional mode of interest was predicted at 401.9 kHz. The other two modes occured at frequencies around 100 kHz (all 3 paddles vibrating in phase) and 160 kHz (side paddles vibrating out of phase and central paddle not moving). Our modelling also showed that this torsion mode of interest was separated from any low quality factor wafer modes by at least 5 kHz, which ensured low contamination to the mechanical loss of the torsion mode.



We experimented with three methods of fabricating our resonators: wet etching using a KOH solution, laser cutting, and DRIE with a mask of SU8 resin. DRIE was chosen as it provided sharper features, compared to tapered walls and other complications (pinholes and undercutting) associated with our efforts with KOH wet etching. Laser cutting was also used and provided a significant improvement over wet etching. However, we observed lower quality factors than for resonators fabricated using DRIE.

The pattern was formed onto a 500 $\mu$m thick silicon boron-doped (100) wafer, DRIE etching through the entire thickness of the wafer. A layer of omnicoat was applied underneath the SU8 2025 resin that was used for photolithography patterning. The omnicoat layer allowed easy removal of the SU8 2025 resin after the DRIE was done. 

The resonator reported here was without optical coating, however we have defined a process by which coating could be put on the sample without covering the torsional rods. This is for future experiments in a cavity designed for 3-mode optoacoustic parametric amplifiers. 

The design presented in this section could lead to excess coupling of the torsional mode to wafer modes. Great care was required to design a suspension system to overcome this problem. This is discussed in the next section.

\section{Experimental Setup and Methods}

%
%
%

\begin{figure}[t]
    \centering
    \includegraphics[width=3.4in]{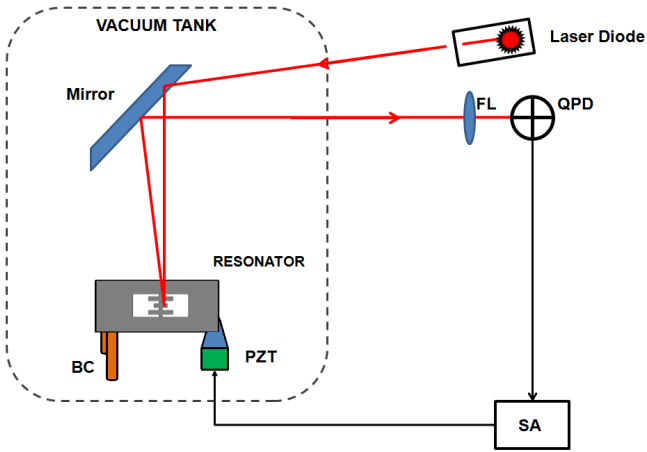}
    \caption{(color online). Schematic drawing of the experimental setup. Brass columns (BC); Piezo-electric transducer (PZT) with a glued-on steel pyramidal tip; Focusing lens (FL); Quadrant photodetector (QPD); Spectrum analyser (SA). Red lines indicate the laser beam path. Black lines indicate the electronic connections. The sample is kept inside a vacuum tank, at room temperature.}
    \label{Schem_Setup}
\end{figure}

The experimental setup used to characterise the fabricated resonator is sketched in Fig. \ref{Schem_Setup}.
The light source was a class III laser diode operating at 26 mW and $\lambda = 650$ nm (red). The laser beam is reflected off a $45\,^{\circ}$ mirror and directed onto the central paddle of the resonator, with a waist size of $\sim$ 300 $\mu$m. The reflected beam is redirected through a focusing lens and onto a quadrant photodetector (QPD).

Clamping the sample at one corner (1 mm by 3 mm area) caused excess losses on the quality factor of the mode of interest. A maximum quality factor value of 9 \texttimes $10^{4}$ was obtained by clamping the wafer at various locations along the edges and corners. Bonding our sample with Yacca gum, a natural resin with low intrinsic loss \cite{Schediwy2005}, also produced quality factors too low for our requirements (6 \texttimes $10^{4}$), as we found the isolation provided by the paddles and torsion rods were insufficient for our required level of quality factors.

We adopted a 3-point support system consisting of two static brass columns of diameter 500 $\mu$m and height 10 mm, and one pyramidal steel tip glued to a piezo-electric transducer (PZT). We chose the 3-point suspension because it allowed a convenient way to excite the resonator by incorporating a PZT into a suspension point. 

As discussed in the previous section, despite the isolation design, there would still be some energy coupling between the torsional mode and some wafer modes. For this reason, the wafer was required to be carefully positioned so that all 3 suspension points were close to a node in the acoustic standing waves on the wafer surface. Finding the correct wafer position could be challenging. However, for the correct position, high quality factors have been observed and are reported in the next section.

The sample was mounted in a vacuum tank as shown in Fig. \ref{Schem_Setup}, in which pressure could be reduced to $10^{-3}$ Pa. The experimental setup was kept at room temperature. A spectrum analyser (SA) generated a driving signal sent to the PZT for excitation of the mechanical modes of the resonator. The QPD measured angular variations in the reflected beam and a corresponding voltage signal was sent to the SA.
%
The driving signals were generated as a fixed sinusoidal or a periodic chirp, both at values between 1 to 40 mV, kept low enough to avoid overdriving, which could cause nonlinear response from the resonator and displacements of the sample. A periodic chirp was used to scan and observe a spectrum of resonant frequencies, while the fixed sinusoidal signal was used to specifically excite a target mode. From the latter, we measured a ringdown curve when the source was switched off. The decay of a ringdown was then fitted with the following formula:

\begin{equation}
\label{DecayRD}
A(t) =  A_{0}e^{\frac{- \pi f t}{Q}} + C_{0},
\end{equation}
where $A(t)$ is the amplitude (in $V_{rms}$) at any time $t$, $A_{0}$ is the amplitude at the time when we switch the source off, $f$ is the mechanical mode frequency (in Hz), $Q$ is the quality factor and the parameter of interest, and $C_{0}$ is the noise floor (in $V_{rms}$), at which the amplitude eventually settles after sufficient time. We repeated the measurement to obtain a value for $Q$ with sufficient accuracy.




\section{Results and Discussion}

\begin{figure}[b]
    \centering
    \includegraphics[width=3.2in]{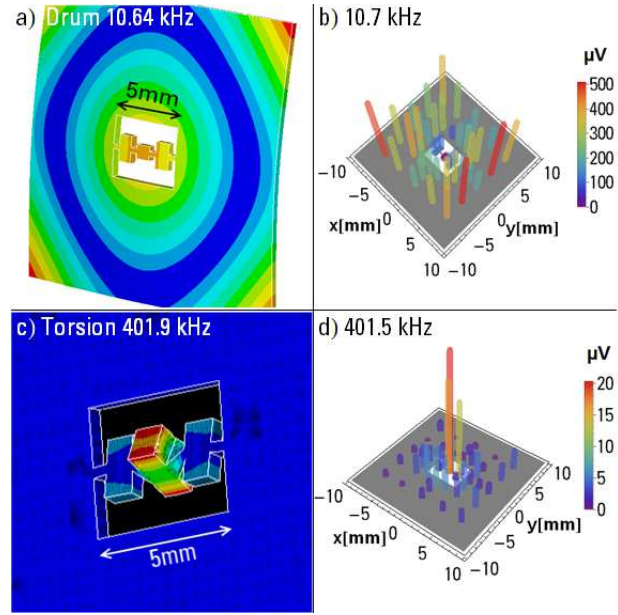}
    \caption{ (color online). Comparison of modelling and measured data for a low and high frequency mode. (a) Fundamental drum mode of the wafer expected at 10.64 kHz. (b) Observed amplitude distribution of angular vibration across the wafer at 10.7 kHz clearly demonstrates the acoustic energy is mostly on the wafer. (c) Torsional mode expected at 401.9 kHz. (d) Observed amplitude distribution of angular vibration at 401.5 kHz, where acoustic energy is concentrated on the central paddle. Measured frequencies agree with modelling predictions to within 400 Hz.
}
    \label{AmplFEM}
\end{figure}

In order to identify the torsional made in which only the central paddle is vibrating, we scanned the wafer surface to measure the angular variations distribution (see Fig. \ref{AmplFEM} (b) and (d)) as a result of PZT excitation. Low frequency modes showed a distribution in which all regions of the wafer are vibrating, while for some high frequency modes, the vibrations are localised on the 3 paddles.

%

We measured the frequencies of the mechanical modes of the resonator between 10 to 500 kHz. We compared these mode frequencies to FEM predictions. Careful measurements were made to obtain correct dimensions of the sample to enable calibration of FEM predictions. The torsion mode of interest and the fundamental drum mode of the wafer were identified experimentally at 401.5 kHz and 10.7 kHz, respectively. This was in good agreement to FEM estimates of 401.9 kHz and 10.64 kHz (Fig. \ref{AmplFEM} (c) and (a)), respectively.

We studied pressure dependence of quality factors and found that acoustic losses from coupling to residual gas became negligible below pressures of 10 Pa. This ensured our measurements at lower pressures were not dominated by gas damping.

\begin{figure}[t]
    \centering
    \includegraphics[width=3.6in]{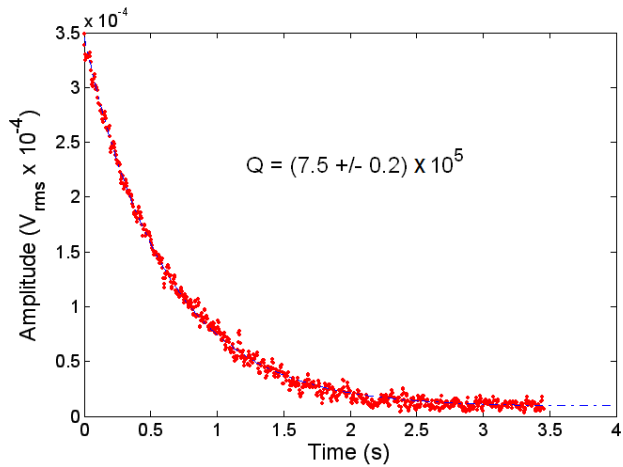}
    \caption{ (color online). Observed quality factor of (7.5$\pm$ 0.2) \texttimes $10^{5}$ at a pressure of $10^{-3}$ Pa and room temperature. A ringdown curve is observed (red data points) when we switch off the fixed sinusoidal signal source at 401.5 kHz. The quality factor is calculated from the exponential fit (blue dashed line) of the ringdown, with error bar obtained from the standard deviation of multiple measurements.}
    \label{HighQ}
\end{figure}
%
%

\begin{table}[b]
    \begin{tabular}{|l|l|l|l|}
       \hline
        Mode & Exp. Freq. (Hz) & FEM Freq. (Hz)& Q       \\ \hline
   Torsion &401 550$\dagger$    &  401 900  & 7.45 \texttimes $10^{5}$ \\ \hline
   Snake &296 220    &  299 800  & 6.56 \texttimes $10^{5}$ \\ \hline
	Twist & 35 500    &   33 450  & 2.16 \texttimes $10^{3}$ \\ \hline
   Drum & 10 700$\dagger$	 &   10 645  & 0.98 \texttimes $10^{3}$ \\ \hline
         
        \hline
    \end{tabular}
\caption{Comparison of the measured frequencies of mechanical modes at room temperature and pressure of $10^{-3}$ Pa to modelling predictions. Both resonator modes (401.5 and 296.2 kHz) have high quality factors ($\sim 10^{5}$), while both wafer modes (35.5 and 10.7 kHz) have low quality factors ($\sim 10^{3}$). $\dagger$ see Fig. \ref{AmplFEM} (a) and (c), for mode shapes.}
\label{tab:Qtable}
\end{table}

%
We obtained an optimal quality factor value of (7.5 $\pm$ 0.2) \texttimes $10^{5}$ for a torsional mode at 401.5 kHz measured at room temperature and pressure of $10^{-3}$ Pa. Ringdown measurements, as shown in Fig. \ref{HighQ}, were repeated to obtain an estimate of the statistical error on the quality factor. This value is higher than quality factors obtained from other silicon structure resonators \cite{Chabot2005,Arcizet2008,Serra2012} presented in Table \ref{tab:ReviewResonators}, and is getting close to quality factor values obtained from silicon bulk samples \cite{Nawrodt2008}.

Using similar methodology, we studied low and high frequency modes and we summarize results for four modes in Table \ref{tab:Qtable}. Both high frequency modes show quality factors significantly higher than for both low frequency modes. When vibrations overlap the locations of the 3-point suspension system, this can create stronger mechanical losses and hence lower quality factor values. Given that both low frequency modes are predicted by FEM to have vibrations on most of the sample area, this could explain why they have quality factors much lower than those of both the high frequency modes, which have vibrations localised on the 3 paddles.

%
%
The `twist' mode presented in Table \ref{tab:Qtable} is a low frequency mode predicted at 35.5 kHz by FEM and observed experimentally at 33.45 kHz. The mode shape is described by considering that any two adjacent corners of the wafer are moving in opposite directions. Both low frequency modes in Table \ref{tab:Qtable} have quality factors in the $10^{3}$ range, low in comparison to higher frequency modes. The second high quality factor mode shown in Table \ref{tab:Qtable} is termed `snake' mode, in which the chain of 3 paddles undergo an S-shaped motion, as predicted by FEM, on a X-Z plane along the Y$=0$ line (see axes on Fig. \ref{AmplFEM} (b) or (d)). This snake mode was predicted at 299.8 kHz and found experimentally at 296.22 kHz. The snake mode also has vibrations localised on the 3 paddles, which explains the relatively high quality factor.

The 3-point suspension system had the disadvantage that it made our results sensitive to small displacements of the sample relative to the 3 points. The quality factors varied greatly as a function of position, reaching a minimum of 40 000 for the torsion mode of interest, for example. Using low frequency and large excitation, we were able to shift the position of the wafer inside the closed tank, and obtain the optimal quality factor.

\section{Conclusion}

This study has shown that micro-mechanical torsional resonators, designed to optimize spatial overlap with the TEM$_{01}$ mode in an optical cavity for 3-mode interactions, can have high mechanical quality factors. 
We have reported a quality factor of (7.5 $\pm$ 0.2) \texttimes $10^{5}$ for a mg-scale resonator at 401.5 kHz and at room temperature. The resonator performance matches the requirements for a proposed 3-mode optoacoustic parametric amplifier. The resonator is designed to ensure low acoustic stress in the optical coating, so as to minimize optical coating losses. We've shown that the resonator can be excited through the substrate without incurring unacceptable suspension losses. FEM predictions were compared to amplitude response across the resonator and the wafer, matching the predicted and measured frequencies and mode shapes. In the future, we propose to assess the use of 3-mode optoacoustic parametric amplifiers to the creation of novel sensor technologies such as electromagnetic sensors, thermal noise sensors or atomic force microscopes. Based on previous cryogenic acoustic loss measurements in bulk silicon \cite{Nawrodt2008}, the resonators presented here, if placed in a cryogenic environment, could be expected to achieve quality factors in the range of $10^{7} - 10^{8}$. Such values would allow a range of experiments in quantum measurements and in advanced transducers to be undertaken. We are currently working with an optically coated resonator in a small optical cavity designed as an optoacoustic parametric amplifier. Results will be reported in a future paper.
%

\begin{acknowledgments}

This work was supported by the Austalian Research Council, the Australian National Fabrication Facility, the French RENATECH network, and a SIRF scholarship from the University of Western Australia. We would like to thank Slawomir Gras and Stefan Danilishin for collaboration on the finite element modelling. 

\end{acknowledgments}


\end{document}